**Switching Chiral Solitons for Algebraic Operation of Topological Quaternary Digits**

Tae-Hwan Kim[1*], Sangmo Cheon[2], and Han Woong Yeom[1,2*]

[1]Department of Physics, Pohang University of Science and Technology (POSTECH), Pohang 37673, Korea. [2]Center for Artificial Low Dimensional Electronic Systems, Institute for Basic Science (IBS), Pohang 37673, Korea.

*Corresponding authors: taehwan@postech.ac.kr, yeom@postech.ac.kr

**Chirality is ubiquitous in nature[1–4] and chiral objects in condensed matter are often excited states protected by system's topology. The use of chiral topological excitations to carry information has been demonstrated, where the information is robust against external perturbations. For instance, reading, writing, and transfer of binary information are demonstrated with chiral topological excitations in magnetic systems, skyrmions[5–12], for spintronic devices[11–17]. However, the next step, the logic or algebraic operation of such topological bits has not been realized yet[18–20]. Here, we show experimentally the switching between solitons of different chirality in a one-dimensional electronic system with $Z_4$ topological symmetry[21,22]. We found that a fast-moving achiral soliton merges with chiral solitons to switch their handedness. This corresponds to the realization of algebraic operation of $Z_4$ topological numbers[23]. Chiral solitons could be exploited for storage and operation of robust topological multi-digit information.**

Topological particle-like excitations can be used as information carriers in next-generation non-volatile memory and logic devices based on their topological robustness. One of the promising candidates for topological information carriers is a magnetic skyrmion or topological magnetic soliton, which is 2D chiral spin configurations[5–12] protected topologically[11–17]. So far, while binary digit memory devices based on skyrmions are successful demonstrated, logic operations have been only conceptually developed[18–20]. The major obstacles are the very sophisticated nanoscale fabrication and the precise material engineering requested for such a device[20].

On the other hand, topological excitations have been known for a long time in 1D electronic systems; a Peierls-distorted atomic chain such as polyacetylene has two topologically distinct ground states, which are connected by a unique topological soliton[24–28]. This topological excitation does not have such chirality as a skyrmion in 2D[22]. However, our team recently discovered new kinds of topological solitons in double Peierls chains of indium atomic wires[21,22]. Double Peierls chains uniquely have topologically distinct four-fold ground states, which are connected by three different topological solitons: one without chirality and chiral ones with different handedness (see Figs. 1a and 1b). They can in principle carry quaternary-digit information with three different topological numbers. Due to their unique $Z_4$ topology, it is natural that $Z_4$ algebraic operations would be possible between these solitons as demonstrated below. Thus, these topological chiral solitons can be utilized not only for topological multi-digit memories in electronic systems but also for logic devices. In contrast to the logic operation proposed with skyrmions, chiral solitons do not need sophisticated engineering since the operation itself is intrinsic to their $Z_4$ topology.

Chirality of solitons can be distinguished by their distinct structures and electronic states[22]. In a wire of double indium zigzag chains, a solitonic phase defect can exist on either of the two chains connecting two degenerate Peierls-dimerized configurations of that chain. This solitonic phase defect in one chain has to interact with the bonding or anti-bonding orbital of the neighbouring chain (Fig. 1a). This interaction breaks the chiral symmetry of the system and provides the chirality to the soliton: left-chiral (LC) and right-chiral (RC) solitons result from the interaction with bonding and anti-bonding orbitals of the neighbouring chain, respectively (Fig. 1a). LC and RC solitons can be distinguished by their scanning tunnelling microscopy (STM) images (inset of Fig. 1b) representing their difference in local charge density configurations. On the other hand, when both chains of a wire have solitonic phase defects on neighbouring sites, the third type of a soliton without chirality emerges, which will be called achiral (AC) soliton hereafter (see the bottom panel of Fig. 1a and the inset of Fig. 1b). These three types of solitons constitute a complete set of a $Z_4$ topological system with four-fold ground states of double Peierls-distorted chains. The distinction of different types of solitons is more striking in their electronic states within the charge-density wave band gap. The electronic states of LC (RC) solitons appear below (above) the Fermi level within the band gap (Fig. 1c) while AC solitons have their states both above and below the Fermi level (the inset of Fig. 2b)[21,22]. The electronic difference results in the fact that LC solitons looks always brighter in STM images than RC ones due to their enhanced density of states in filled states (Supplementary Fig. 1).

Topological solitons are highly mobile but can be trapped by other defects or impurities to allow the STM observation[21]. Spontaneously charged solitons[22] would be easily trapped by a defect charged oppositely through the Coulomb attraction. The Coulomb interaction would shrink the size of a trapped soliton significantly[27] as also observed for the present solitons trapped (see the inset of Fig. 1b for unbound chiral solitons). The characteristic electronic state of a trapped soliton[22] remain intact as it is endowed from the topology of the system (Fig. 1c).

During long-term STM observations of a trapped chiral soliton, it changes its handedness abruptly and occasionally (Fig. 1b). This occasional but very fast switching is not fully captured by repeated STM imaging with a 50-sec scan interval. However, a scanning tunnelling spectroscopy (STS) measurement at a fixed location repeated every 3 sec can track switching events better due to the unique electronic states of AC, RC, and LC solitons[22]. The temporal resolution for an event changing a particular electronic state can be as fast as 0.01 sec in this measurement, ~5,000 times faster than that of sequential STM imaging. As shown in Supplementary Fig. 2, for a particular series of four sequential STS voltage sweeps on a trapped soliton, the density of states (the intensity of the normalized $dI/dV$ signal) changes abruptly twice (indicated by red arrows). By comparing the STS spectra before and after these events with the reference spectra of chiral solitons, we can identify that the initial RC soliton is switched to the LC one ($t$ = 3.1 s) and back to the RC one ($t$ = 5.4 s). For an extended STS observation on a single chiral soliton, we can observe many telegraph switchings between the states above and below the Fermi level (Fig. 1d and Fig. 2a). Each of these switchings in electronic states corresponds to the chirality switching of the soliton (Fig. 2b) since the in-gap states have one-to-one correspondence with chirality of solitons. In addition, for a few events, one can find STS intensity for both filled and empty states (see Figs. 2a, 2b and the inset of 2b), which are unambiguously identified as the presence of a AC soliton.

In order to understand the mechanism of the chirality switching of solitons, we note on the fact that each switching event is accompanied by the hopping of trapping defects. As shown in Fig. 1b, one of the two trapping defects (the right one) hops along the chain by an integer multiple of the lattice unit cell ($a_0$ = 0.384 nm). Figures 2c and 2d indicate with no doubt that the hoppings of this defect are uniquely related to the handedness of chiral solitons. That is, a $1a_0$-hopping changes the chirality of the trapped soliton while a $2a_0$-hopping does not without any exception. Since the periodicity of the Peierls chain is $2a_0$, the $1a_0$-hopping of defects inevitably changes nearby ground states so that it must involve a soliton in contrast to a $2a_0$-hopping maintaining the same ground states before/after the hopping. Indeed, our previous work established that the $1a_0$-hopping is induced by an AC soliton passing through defects[21]. Therefore, we can conclude that a fast moving AC soliton giving rise to the $1a_0$-hopping is responsible for switching the handedness of a chiral soliton.

This fast moving AC soliton can occasionally be captured in the experiment to support the above explanation. As mentioned above for Fig. 2b, rather rarely, both filled and empty soliton states are similarly high at the same time (marked with red ovals in Fig. 2b), which correspond to the STS spectra of AC solitons (the inset of Fig. 2b)[22]. This finding indicates unambiguously that AC solitons momentarily exist as required for $1a_0$-hoping of defects and switching the handedness of chiral soliton.

We show further below that the AC soliton is indeed topologically requested to switch the chirality of the soliton. Figure 3a shows the schematic diagrams describing the chiral switching of a RC soliton induced by a left-going AC soliton. For simplicity, we ignore the trapping defects (see Supplementary Fig. 3 when including the trapping defects). Initially, both the RC and AC solitons are away and bridge different ground states. When they come close enough to each other, the ground state between two solitons disappears and two solitons together become equivalent to a LC soliton in terms of a boundary condition. Thus, they can eventually merge to create a LC soliton due to energetic reasons: one soliton has a lower energy than two (Fig. 3b and Supplementary Fig. 4). More detailed energetics will be considered below. If another AC soliton comes again, the LC soliton would switch back to a RC soliton by the same procedure.

This chiral switching is in fact intrinsic to the topology of the present system. The system's topology can by written in the topological order-parameter space (Fig. 3c) with the order parameters $m_x$ and $m_z$ representing dimerization sublattice and interchain coupling potential terms, respectively[22] and the four-fold ground states denoted as **Aa**, **Ab**, **Ba**, and **Bb**. For the particular chiral switching in Fig. 3a, **Aa**→**Ba** (RC) soliton and **Ba**→**Ab** (AC) soliton merge into **Aa**→**Ab** (LC) soliton. In another case, two RC solitons can merge into an AC soliton: **Aa**→**Ba** + **Ba**→**Bb** = **Aa**→**Bb** (Fig. 3c). We can expand this consideration (Fig. 3c) to the cyclic group under addition modulo four with four elements of LC, RC, and AC soliton as well as an identity element (denoted as G in Fig. 3d indicating the ground state without a soliton). This cyclic group of the present system is the same as $Z_4$ abelian four-element group[23] and the soliton switching corresponds to algebraic operation of $Z_4$ topological numbers (Supplementary Fig. 5).

We can elaborate the energetics of the merging of two solitons. Self-consistent tight-binding calculations based on generalized Su-Schrieffer-Heeger Hamiltonian[22] are utilized to estimate soliton formation energies. Neutral AC solitons are the lowest topological excitations while RC and LC solitons prefer to be positively and negatively charged, respectively, when they are created (see Supplementary Fig. 4). In the case of LC+AC→RC operation, a negatively charged LC soliton and a neutral AC soliton should merge into a negatively charged RC soliton under the charge conservation. However, a negatively charged RC soliton has a much higher formation energy and the charge-conserved merging is energetically unfavourable (Fig. 3b). Only when the merged RC soliton becomes positively charged possibly through the charge transfer from/to nearby defects, a merged soliton can have an energy gain over the energy sum of two initial solitons. This charge transfer with defects is not directly probed here but plausible since the Coulomb interaction between charged solitons and defects is prerequisite for the soliton trapping observed. The charge transfer is further hinted by the observation that apparent STM heights of defects systemically change depending on the chirality of solitons (Fig. 1b and Supplementary Fig. 6).

We demonstrate that two solitons with different chirality can merge into another chiral soliton. For normal solitons and most of other topological excitations observed, the algebraic operations are rather trivial as annihilation of two excitations due to their simpler topology of $Z_2$. The realization of non-trivial operation of topological excitations yielding a different soliton is unprecedented and unique to the $Z_4$ topology. The switching between three different types of solitons allows us to use them as quaternary-digit information carriers as well as logic operators. This will open a door into logic devices using topological excitations and more generally into a new field of electronics called 'solitonics'.

**Methods**

The STM/STS experiments were performed in an ultrahigh-vacuum (8 x$10^{-9}$ Pa) cryogenic STM (T = 78.150 K). The sample temperature was controlled to a precision of at least ± 1 mK by using a temperature controller in order to minimize unwanted thermal drift during measurements. Sample preparation procedures are described elsewhere[21,29]. All STM images were recorded in the constant-current mode with electrochemically etched W tip. The sample bias and tunnelling current were set to -0.5 V and 0.1 nA, respectively for STM/STS measurements. STS measurements were performed by recording the *dI/dV* spectra via a lock-in technique with feedback open using a modulation frequency of about 1 kHz at a root-mean-square voltage of 7 mV.

Self-consistent tight-binding calculations were performed with generalized Su-Schrieffer-Heeger Hamiltonian[22] to estimate soliton formation energies and to verify the energetic preference of soliton operations. Soliton states were calculated for a double chain of more than 800 atomic sites per chain with the periodic boundary condition. Soliton state spectra and energy change were obtained as a function of the distance between two solitons via adiabatic approach.

**Acknowledgements** We thank S.-H. Lee for discussion in the early stage and the technical help on the calculation method. This work was supported by IBS-R014-D1. T.-H.K. was also supported by Basic Science Research Program (Grant No. NRF-2014R1A1A1002205) and the SRC Center for Topological Matter (Grant No. 2011-0030046) through the National Research Foundation (NRF) of Korea funded by the Ministry of Science, ICT & Future Planning.

**Author Contributions** T.-H.K. performed STM/STS measurements; S.C. performed tight-binding calculations; T.-H.K. and H.W.Y. analysed data and wrote the paper. All authors discussed the results and commented on the manuscript.

**Competing financial interests** The authors declare no competing financial interests.

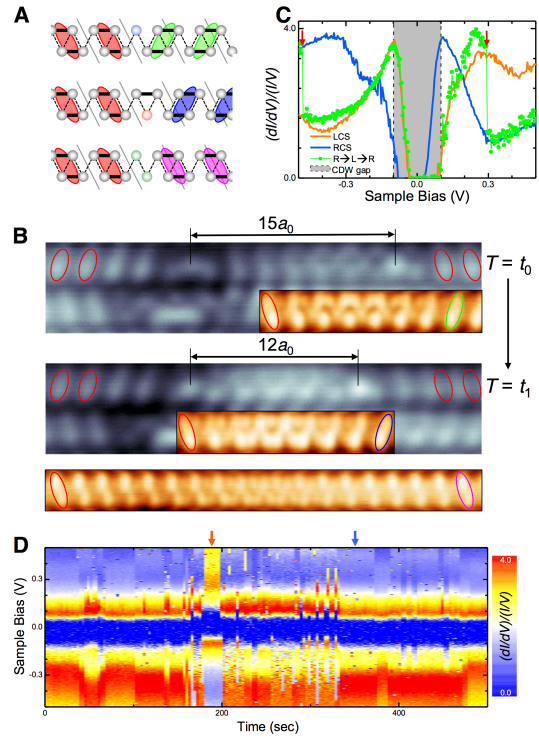

**Figure 1| Characterizing chiral switching between topological solitons. a**, Schematics of right-chiral (RC), left-chiral (LC), and achiral (AC) solitons from top. Coloured ovals represent four-fold ground states with different topology. For simplicity, the size of solitons is reduced. Thick solid and dashed lines between atoms indicate Peierls dimers and interchain coupling, respectively. **b**, Sequential STM imaging reveals chiral switching between RC ($T = t_0$) and LC ($T = t_1$) solitons is related to the distance between two defects holding chiral solitons. Red ovals clearly exhibit that the ground state on the right side of the defects is also altered after the chiral switching. Insets are typical STM images of unbound chiral solitons keeping the same lateral position as trapped ones as well as an unbound achiral soliton (bottom). Each coloured oval represents the four-fold ground states as the same colour codes in **a**. **c**, Green curve shows chiral switching events from RC to LC and from LC to RC indicated by red arrows. The charge-density wave (CDW) gap is highlighted by the shaded regions. **d**, Time-traced STS spectra clearly show two dominant soliton states indicated by orange (LC) and blue (RC) arrows as well as several abrupt switching events between LC and RC soliton states.

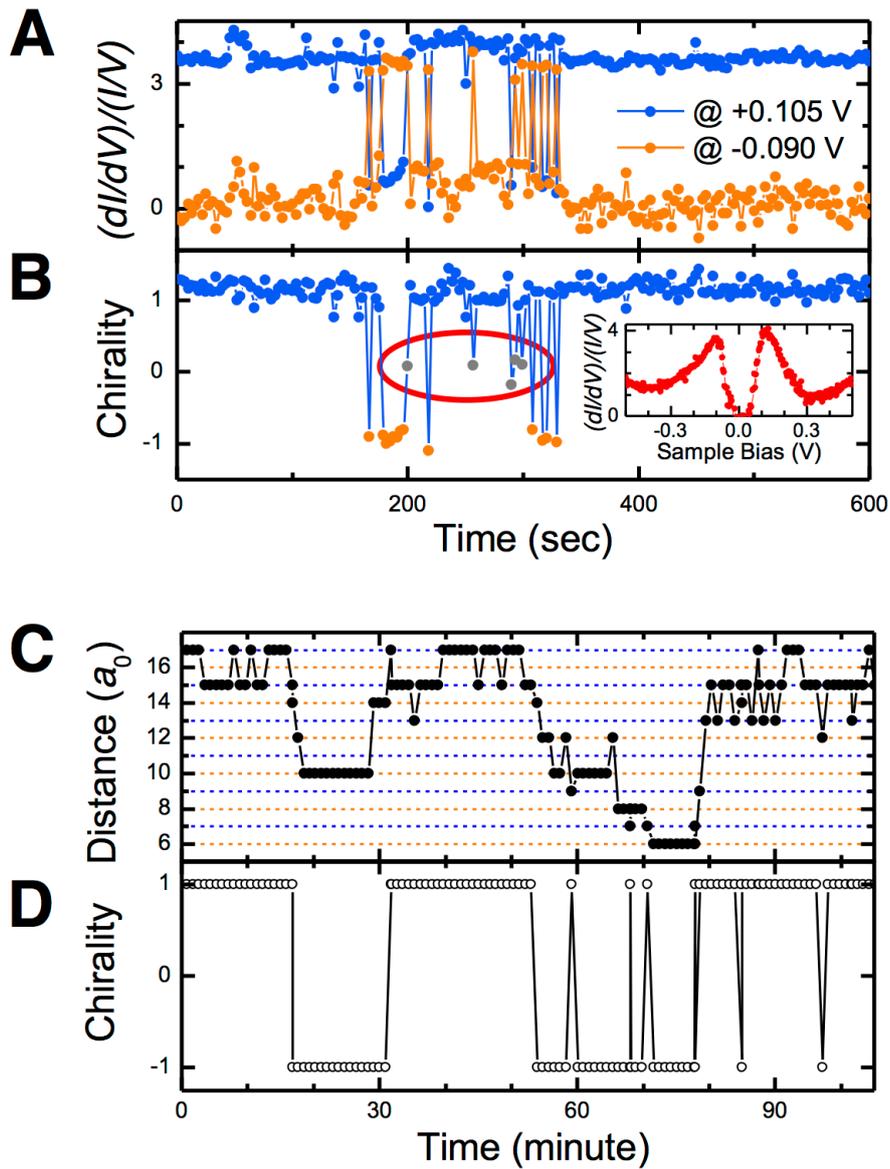

**Figure 2| Trace section of the time-dependent chiral switching. a**, Time-dependent $dI/dV$ signal of soliton states collected at +0.105 V (blue dots for RC, $I_R$) and -0.090 V (orange dots for LC, $I_L$) from Fig. 1e. **b**, Time-dependent chirality defined by asymmetry of $dI/dV$ signals, $A=(I_R-I_L)/(I_R+I_L)$. A red oval indicates intermediate states between LC and RC. Inset shows STS spectrum obtained at one of the intermediate states, which clearly identifies an AC soliton. **c**, Time-dependent distance between two defects trapping chiral solitons drawn from consecutive STM imaging (120 images). **d**, Time-dependent chirality obtained from the polarity of the defect distance.

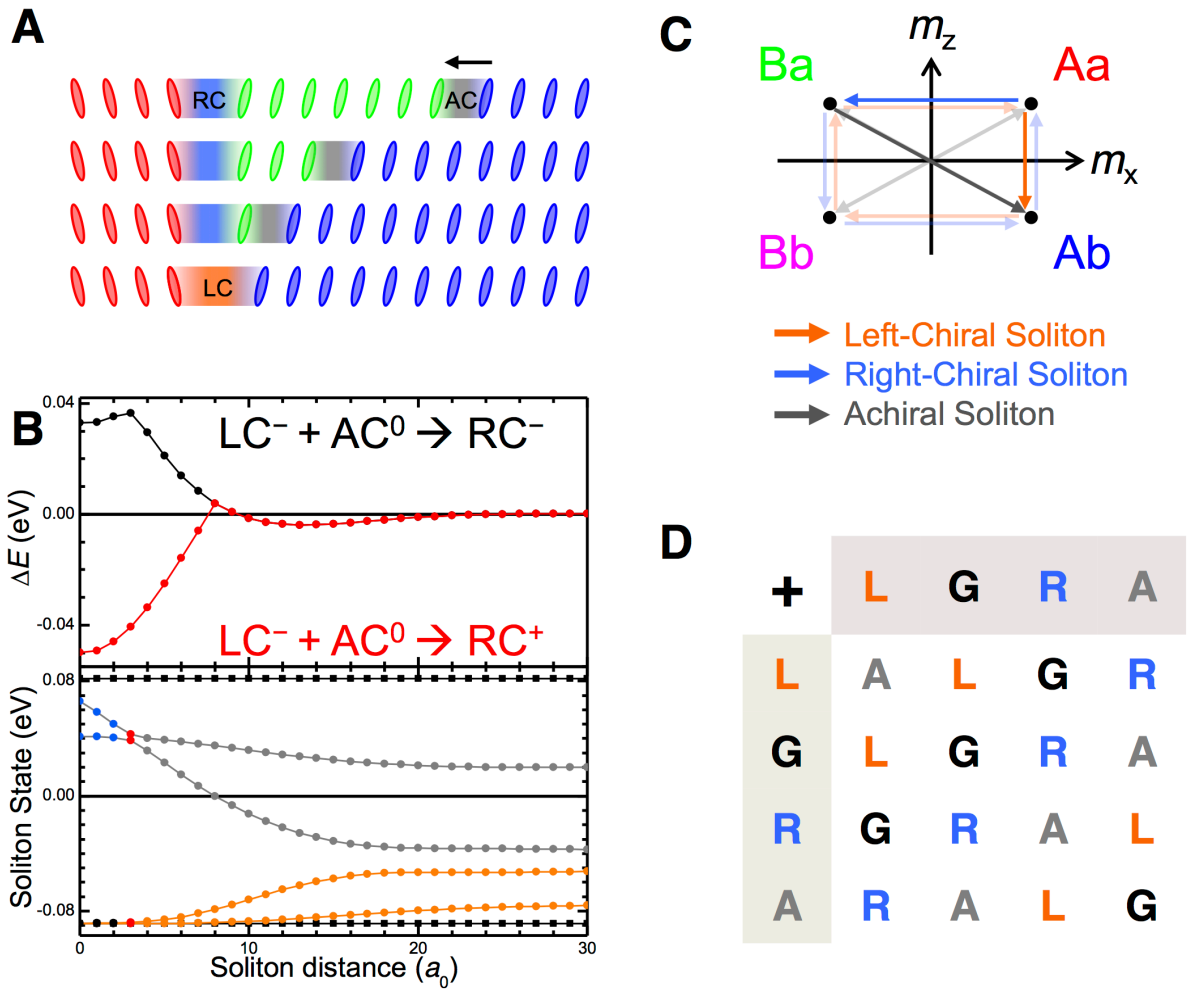

**Figure 3| Algebraic operation using solitons. a**, Schematics showing chiral switching from RC to LC by an AC soliton. Coloured ovals represent four-fold ground states. **b**, Total energy change as a function of a distance between LC and AC (upper panel). Soliton states evolution depending on a distance between LC and AC (lower panel). Less than the critical distance ($3a_0$), AC soliton states are switched to RC soliton states while LC soliton states disappear. **c**, Algebraic operation of solitons (RC+AC=LC is highlighted) in the order-parameter space. **Aa**, **Ab**, **Ba**, and **Bb** indicate four-fold degenerate ground states. Blue, orange, grey arrows represent RC, LC, AC solitons, respectively. **d**, Algebraic operation table showing $Z_4$ abelian group. G, A, L, and R indicate ground states, AC, LC, and RC solitons, respectively.